\documentclass[twocolumn,preprint]{emulateapj-rtx4}
\pdfoutput=1 
\usepackage{graphicx}
\usepackage{amsmath,amssymb}
\usepackage{txfonts}
\usepackage{epstopdf}
\usepackage{mathrsfs}
\usepackage{color}
\usepackage{multirow}

\shorttitle{Reconnection as a driver for a sub-ion scale cascade}
\shortauthors{Franci et al.}

\newcommand\Bv{{\bf B}}
\newcommand\Jv{{\bf J}}

\newcommand\di{d_{\rm i}}

\begin{document}

\title{Magnetic reconnection as a driver for a sub-ion scale cascade in plasma turbulence}

\author{Luca~Franci}
\affiliation{Department of Physics and Astronomy, University of Florence, Via G. Sansone 1, I-50019 Sesto F.no, Firenze, Italy}
\author{Silvio~Sergio~Cerri}
\affiliation{Physics Department ``E. Fermi€'', University of Pisa, Largo B. Pontecorvo 3, I-56127 Pisa, Italy}
\author{Francesco~Califano}
\affiliation{Physics Department ``E. Fermi€'', University of Pisa, Largo B. Pontecorvo 3, I-56127 Pisa, Italy}
\author{Simone~Landi}
\affiliation{Department of Physics and Astronomy, University of Florence, Via G. Sansone 1, I-50019 Sesto F.no, Firenze, Italy}
\author{Emanuele~Papini}
\affiliation{Department of Physics and Astronomy, University of Florence, Via G. Sansone 1, I-50019 Sesto F.no, Firenze, Italy}
\author{Andrea~Verdini}
\affiliation{Department of Physics and Astronomy, University of Florence, Via G. Sansone 1, I-50019 Sesto F.no, Firenze, Italy}
\author{Lorenzo~Matteini}
\affiliation{Department of Physics, Imperial College London, London SW7 2AZ, UK}
\author{Frank~Jenko}
\affiliation{Max Planck Institute for Plasma Physics, Boltzmannstr. 2, 85748 Garching, Germany}
\author{Petr~Hellinger}
\affiliation{Astronomical Institute, CAS, Bocni II/1401, CZ-14100 Prague, Czech Republic}

\begin{abstract}
A new path for the generation of a sub-ion scale cascade in
collisionless space and astrophysical plasma turbulence, triggered by
magnetic reconnection, is uncovered by means of high-resolution
two-dimensional hybrid-kinetic simulations employing two complementary
approaches, Lagrangian and Eulerian, and different driving
mechanisms. The simulation results provide clear numerical evidences
that the development of power-law energy spectra below the so-called
ion break occurs as soon as the first magnetic reconnection events
take place, regardless of the actual state of the turbulent cascade at
MHD scales. In both simulations, the reconnection-mediated small-scale
energy spectrum of parallel magnetic fluctuations exhibits a very
stable spectral slope of $\sim -2.8$, whether or not a large-scale
turbulent cascade has already fully developed. Once a quasi-stationary
turbulent state is achieved, the spectrum of the total magnetic
fluctuations settles towards a spectral index of $-5/3$ in the MHD
range and of $\sim -3$ at sub-ion scales.
\end{abstract}



\section{Introduction}

Turbulent dynamics and its interplay with magnetic reconnection in
collisionless plasmas is of great interest in many different
astrophysical environments, e.g., in the interstellar medium, in
accretions disks, and in stellar coronae and winds.  Direct in-situ
measurements of near-Earth turbulent plasmas, such as the solar wind
(SW) and the terrestrial magnetosheath, have led to increasingly
accurate constraints on the turbulent energy spectra and on the
magnetic field structure~\citep{BrunoCarboneLRSP2013,StawarzJGR2016,
  MatteiniMNRAS2017}.  These observations determine the typical
spectral slopes for turbulent electromagnetic fluctuations and reveal
the presence of a break in the power spectra at ion kinetic
scales~\citep{BalePRL2005, AlexandrovaPRL2009, ChenPRL2010,
  SahraouiPRL2010}, separating a magnetohydrodynamic (MHD) inertial
cascade from a kinetic cascade.  The former is generally characterized
by a $-5/3$ slope in the magnetic power spectrum, whereas the latter
is quite steeper, with a spectral index around $\sim-2.8$.  The
typical picture of the full cascade assumes an energy transfer
towards small scales mainly made by quasi-2D Alfv\'enic fluctuations
in the MHD range \citep{MatthaeusJGR1982, BieberJGR1996} and by a
mixture of dispersive modes in the ion kinetic range
\citep{StawickiJGR2001, GaltierPOP2003, HowesJGR2008, SchekochihinAPJS2009, BoldyrevAPJL2012,
  BoldyrevAPJ2013}, corresponding to
local nonlinearities in Fourier space. However, 
embedded in this dynamics is the interaction of coherent
structures where nonlinear interactions are rather local in real
space: vortices, current sheets, magnetic and flow shears, are seen as
birthplace of the intermittent behavior of the turbulence where
``dissipation" is thought to be partially, but not completely,
localized~\citep{ZhdankinAPJ2013, OsmanPRL2014, WanPRL2015,
  ServidioJPP2015, NavarroPRL2016}. The disruption of current sheets
via magnetic reconnection is very efficient in accelerating particles,
creating coherent structures and electromagnetic fluctuations at
ion-scales~\citep{MaJGR1999, SturrockAPJ1999, LoureiroPRL2013,
  GrecoApJ2016}, which allow for and/or enhance the nonlinear transfer
of energy around and below the ion scales~\citep{CerriNJP2017}, in a
way similar to what happens when the plasma is driven toward kinetic
instabilities~\citep{HellingerApJ2015,HellingerApJ2017}. For these
reasons, the interpretation of the turbulent cascade solely in terms
of linear modes is problematic and
unsatisfactory~\citep{MatthaeusAPJ2014}. In particular, we believe
that coherent structures do play an active role in characterizing the
turbulent path.

The study on turbulent reconnection dates back to the seminal work of
\citet{Matthaeus_Lamkin_1986}, and mainly focused on
magnetohydrodynamics aspects \citep[e.g.][]{Lazarian_Vishniac_2009,
  Lapenta_Bettarini_2011, Servidio_al_2011, Eyink_2015,
  Lazarian_al_2015, Boldyrev_Loureiro_2017, Mallet_al_2017}. Only
recently, the improved numerical resources and techniques allowed
studying the interplay between turbulence and reconnection in
collisionless plasma~\citep[e.g.][]{Burgess_al_2016, CerriNJP2017, Pucci_al_2017}.

In this Letter, by means of high-resolution kinetic-hybrid Lagrangian
and Eulerian simulations, we provide numerical evidences that magnetic
reconnection can act as a driver for the onset of the sub-ion turbulent
cascade. Following the formation of the turbulent
spectrum, we show that the power-law kinetic spectrum is formed as
soon as magnetic reconnection starts occurring in current sheets,
independently from the existence of a fully developed spectrum at
MHD scales. Such result does not depend on the numerical
approach and on the method adopted to drive the turbulent dynamics
(forced or decaying turbulence).

We have reasons to believe that, once the sub-ion spectrum is settled
down and reaches a stationary power-law regime, reconnection still
remains an important energy channel feeding the small-scale
turbulence. Although we cannot quantitatively evaluate the competition
between reconnection and the standard wave-wave interaction energy
transfer mechanism, we discuss elements in favor of our conjecture.

\section{Simulations setup} 

Our model integrates the Vlasov-Maxwell equations in the hybrid
approximation, where fully-kinetic ions are coupled to a neutralizing
massless electron background and quasi-neutrality is assumed~\citep{WinskeSSRv1985, MatthewsJCP1994,
  ValentiniJCP2007}.  We present two direct numerical simulations
employing different approaches, both in the numerical method
used to integrate the Vlasov equation and in the way to achieve
the turbulent state: (i) freely-decaying fluctuations with the 
Lagrangian hybrid particle-in-cell (HPIC) code CAMELIA and 
ii) continuously-driven fluctuations by an external low-amplitude
forcing with the Eulerian hybrid Vlasov-Maxwell (HVM) code.  
In both codes, the ion
inertial length, $d_i$, and the inverse ion gyro-frequency,
$\Omega_i^{-1}$, are used as the characteristic spatial and temporal
units, respectively. Both simulations are ``2.5D'', with a uniform
mean magnetic field perpendicular to the simulation plane.

We consider the same plasma beta for ions and electrons,
$\beta_{\rm i} = \beta_{\rm e} = 1$, with isothermal electrons and
no initial ion temperature anisotropy.  In both simulations, the
energy-containing scales ($k_\perp d_{\textrm{i}} \lesssim0.3$) and
the scales significantly affected by numerical effects ($k_\perp
d_{\textrm{i}} \gtrsim 10$) are basically the same. The HPIC 
simulation employs freely-decaying, large-amplitude 
initial magnetic and velocity perturbations, purely
perpendicular to the mean magnetic field~\citep{FranciAPJL2015, FranciAPJ2015}. 
The HVM simulation employs instead a
3D small-amplitude initial magnetic perturbation with no velocity
counterpart, fed by a continuous external injection of compressible
fluctuations~\citep{CerriAPJL2016}. The grid size is $256 \,\di$ 
for the HPIC and $20 \pi\,\di$ for the HVM with $2048^2$ and $1024^2$ uniformly 
distributed grid points, respectively. The HPIC run employs 64000 particles-per-cell,
while the HVM run employs a $51^3$ points in the velocity domain.
Energy accumulation at the smallest scales is prevented by a fine-tuned explicit
resistivity in the HPIC and by numerical filters in the HVM.
For further details on the two numerical methods and the initial
conditions see~\citet{CerriJPP2017}. 

\begin{figure}[t]
\includegraphics[width=0.49\textwidth]{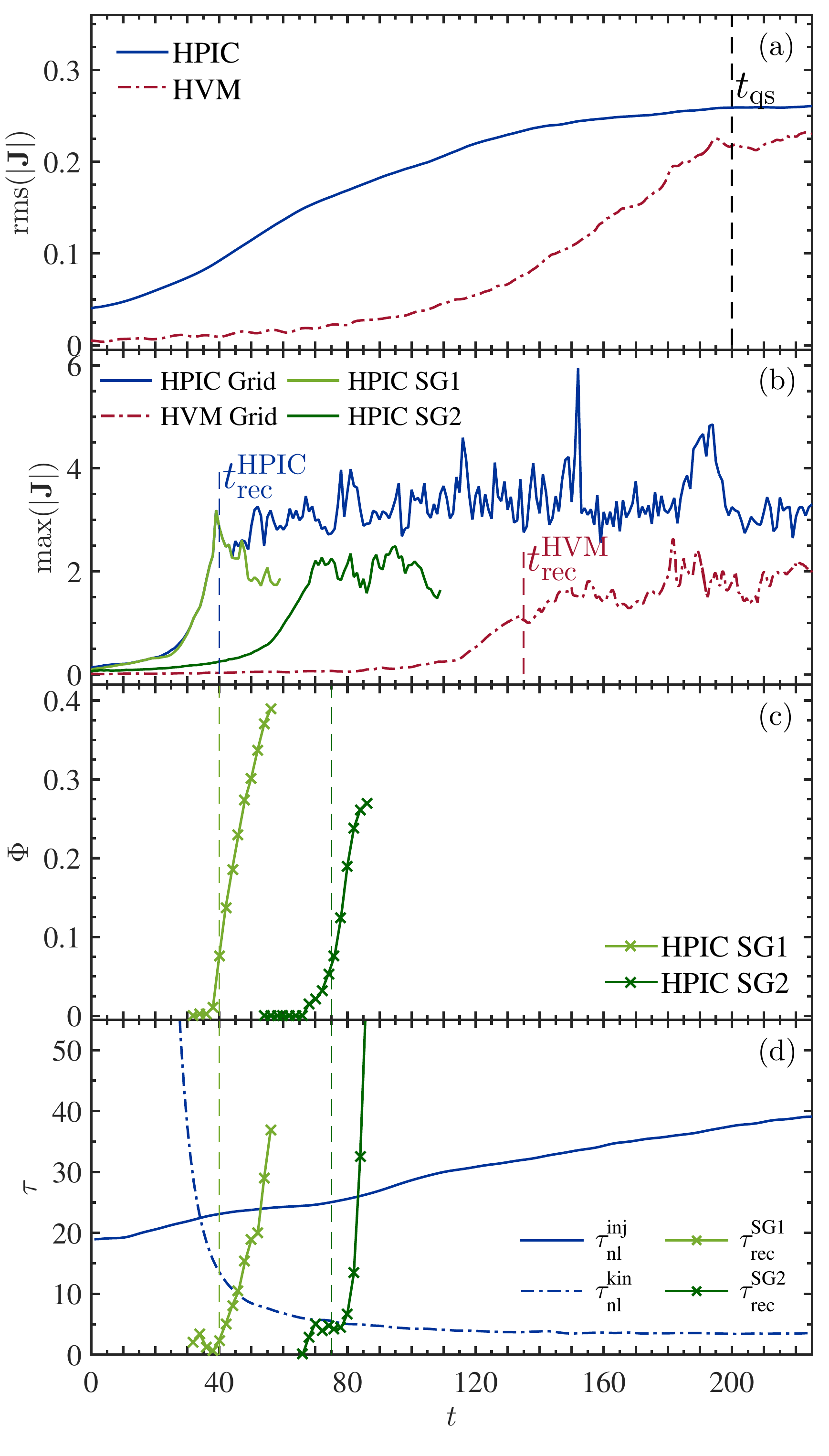}
\caption{Time evolution of a few global and local quantities. 
  \textit{Panel a}: $\mathrm{rms}(|\Jv|)$ in
  the HPIC (blue) and the HVM (red) runs. The black line marks the
  quasy-steady state at $t_{\mathrm{qs}} \sim 200$. \textit{Panel b}:
  $\max(|\Jv|)$ in the whole HPIC (blue) and HVM
  (red) grids, and in two HPIC sub-grids (cf. Fig.~\ref{fig:2DPlots}
  c-f).  The blue and red vertical lines mark the time when magnetic
  reconnection start occurring, $t_{\mathrm{rec}}^{\mathrm{HPIC}}$ and
  $t_{\mathrm{rec}}^{\mathrm{HVM}}$, respectively. \textit{Panel c}:
  Reconnected flux, $\Phi$, in the two HPIC grids. \textit{Panel d}:
  comparison between the eddy turnover time at the injection scale,
  $\tau^\mathrm{inj}_\mathrm{nl}$, and at kinetic scale,
  $\tau^\mathrm{kin}_\mathrm{nl}$, and the inverse reconnection 
  rate in the two HPIC grids, $\tau^{\mathrm{SG1}}_\mathrm{rec}$
  and $\tau^{\mathrm{SG2}}_\mathrm{rec}$.}
\label{fig:TimeEvolution_RmsMaxJ}
\end{figure}
\begin{figure*}
\includegraphics[width=0.9\textwidth]{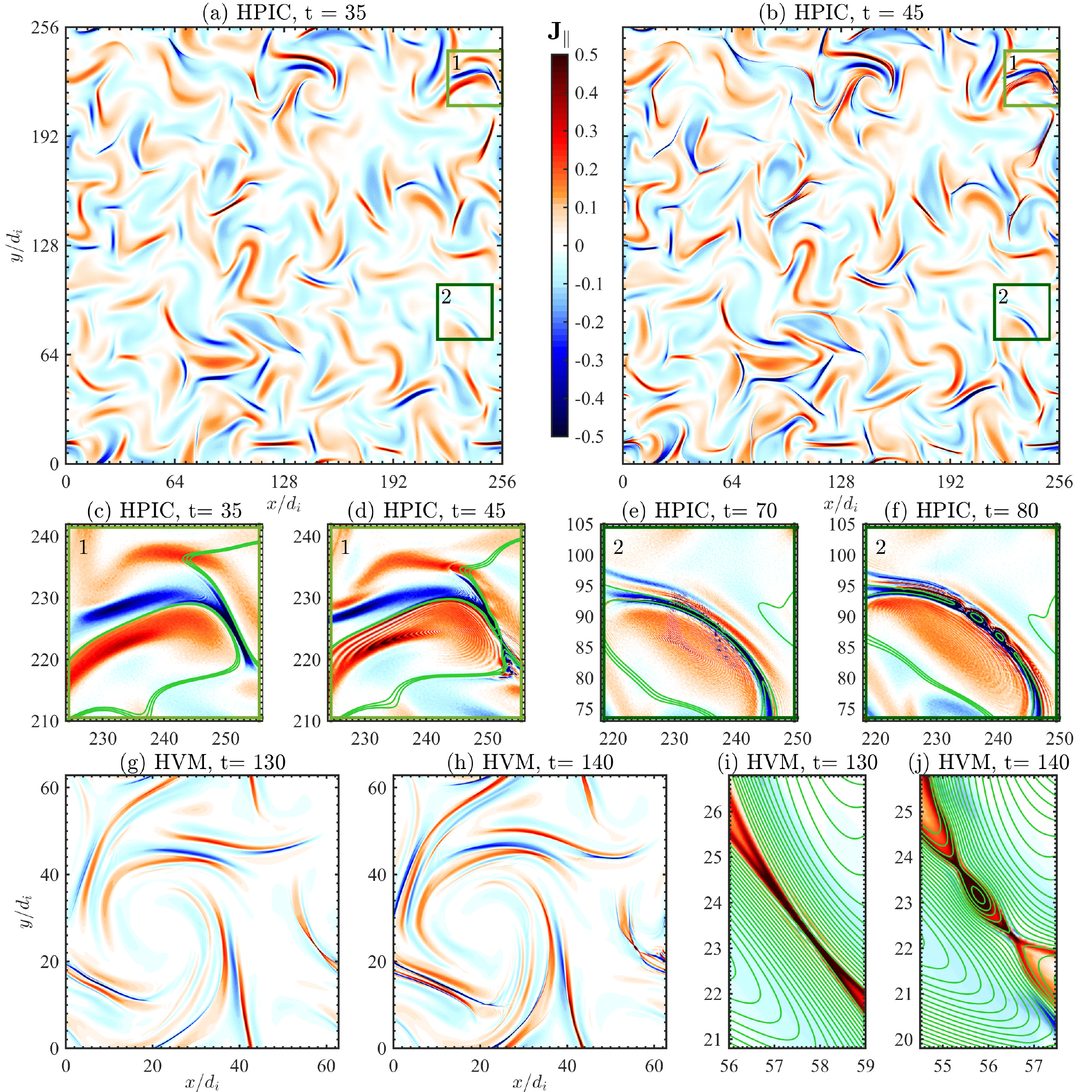}
\caption{Contours of the out-of-plane current density,
  $\Jv_{\parallel}$, before and after the onset of magnetic
  reconnection. \textit{Panels a-b}: whole HPIC
  grid around $t_{\mathrm{rec}}^{\mathrm{HPIC}} \sim
  40$. \textit{Panels c-f}: two HPIC sub-grids, containing
  one reconnecting current sheet. Additionally, isocontours of 
  $A_{||}$ are drawn in green. \textit{Panels g-j}: 
  the same for the HVM simulation, around
  $t_{\mathrm{rec}}^{\mathrm{HVM}} \sim 135$.}
\label{fig:2DPlots}
\end{figure*}

\section{Results} 

As outlined by early MHD and, more recently, by
kinetic simulations, an intrinsic feature of magnetized plasma
turbulence is the formation of current sheets between large-scale
eddies and their subsequent disruption via magnetic reconnection,
generating a variety of small-scale structures and
fluctuations~\citep{Matthaeus_Lamkin_1986, BiskampBOOK2003, 
  KarimabadiPOP2013, FranciAPJ2015,CerriNJP2017}.  The
root-mean-square (rms) value of the current density, $|\Jv|$, represents
a good marker of the turbulent activity~\citep{MininniPRE2009}.  
The time evolution of $\mathrm{rms}(|\Jv|)$ 
(Fig.~\ref{fig:TimeEvolution_RmsMaxJ}a) is quite different in the
two simulations at early times, due to the
different initial conditions: in HPIC, the
relatively large initial fluctuations rapidly drive the system toward
a strong turbulent regime and generate rapidly many current sheets, 
resulting in an increase of $\mathrm{rms}(|\Jv|)$. In HVM, the turbulent 
dynamics is reached later, thanks to the continuous injection of 
momentum, but still $\mathrm{rms}(|\Jv|)$
starts to grow when the current sheet formation
phase begins. In both cases, however, the growth saturates 
at $t_{\mathrm{qs}} \sim 200$, followed by a plateau.
The time evolution of the maximum of $|\Jv|$
(Fig~\ref{fig:TimeEvolution_RmsMaxJ}b) is
coherent with the rms time history and can be used as a proxy for
reconnection events. In both simulations, $\max(|\Jv|)$ is very small 
at early times and then rapidly increases, exhibiting a series of
peaks; such maxima correspond to the evolution of 
current sheets which, once formed, shrink down toward a critical width
of the order of the ion inertial length~\citep{FranciAPJ2016} before they start to reconnect, generating
chains of magnetic islands (O-points)~\citep{CerriNJP2017}, and locally reducing the current density intensity.
To provide further evidence of this mechanism, we compute the local
maxima in two HPIC sub-grids,  
where only one intense current sheet is present. 
The corresponding evolution of $\max(|\Jv|)$ 
confirms what expected: $\max(|\Jv|)$ is initially
very small and then quickly increases,
reaching a local maximum after which it suddenly relaxes.  
Moreover, Fig~\ref{fig:TimeEvolution_RmsMaxJ}c shows the reconnected
flux, i.e., the difference between the out-of-plane vector potential
$A_{||}$ at one of the O-points and its nearest X-point, $\Phi =
A_{||}^\mathrm{O} - A_{||}^\mathrm{X}$, for the two HPIC subgrids mentioned
above. In both cases, $\Phi$ increases very rapidly just before the
local maxima of $\max(|\Jv|)$.  Based on such analysis, we define
$t_{\mathrm{rec}}^{\mathrm{HPIC}} \sim 40$ as the time from which
reconnection is dynamically active in the HPIC case, and similarly
$t_{\mathrm{rec}}^{\mathrm{HVM}} \sim 135$ for the HVM case.  Note
that $t_{\mathrm{rec}}^{\mathrm{HPIC}}$ and
$t_{\mathrm{rec}}^{\mathrm{HVM}}$ are comparable with the initial
eddy turnover time at the injection scale, which is 
$\tau^{\mathrm{inj}}_\mathrm{nl} \sim 20$
and $\sim 120$ for HPIC and HVM, respectively.
This is compatible with the fact that
the formation and shrinking of the first current sheets is
due to the dynamics of the largest-scale eddies.
Such nonlinear times have been estimated as 
$\tau^{\mathrm{inj}}_\mathrm{nl}= 
(k_{\perp}^\mathrm{inj} \mathbf{u}_i^\mathrm{inj})^{-1}                            
\approx (k_{\perp}^\mathrm{inj} \Bv^\mathrm{inj})^{-1}$,
where the ion bulk velocity fluctuations, $\mathbf{u}_i$,
and the magnetic fluctuations, $\Bv$, have been evaluated at
$k_{\perp}^\mathrm{inj} \, d_i = 0.15$. 

A qualitative view of the current sheet disruption is obtained by
comparing the out-of-plane current density, $\Jv_\|$, before and after
the first reconnection events occur in the whole HPIC grid
(Fig.~\ref{fig:2DPlots}a-b).  The only difference is that some current
sheets have shrunk and grown in intensity and reconnection has
occurred somewhere, generating X-points and magnetic islands, without
significant changes at large scales. This process is highlighted by
focusing on local changes in $\Jv_\|$ and in the isocontours of 
$A_{||}$ in correspondence with an early
(Fig.~\ref{fig:2DPlots}c-d, $t=[35,~45]$) and a late reconnection
event (e-f, $t=[70,~80]$) corresponding to local maxima of
$\max(|\Jv|)$ (cf. Fig.~\ref{fig:TimeEvolution_RmsMaxJ}b).

\begin{figure}[t!]
\includegraphics[width=0.47\textwidth]{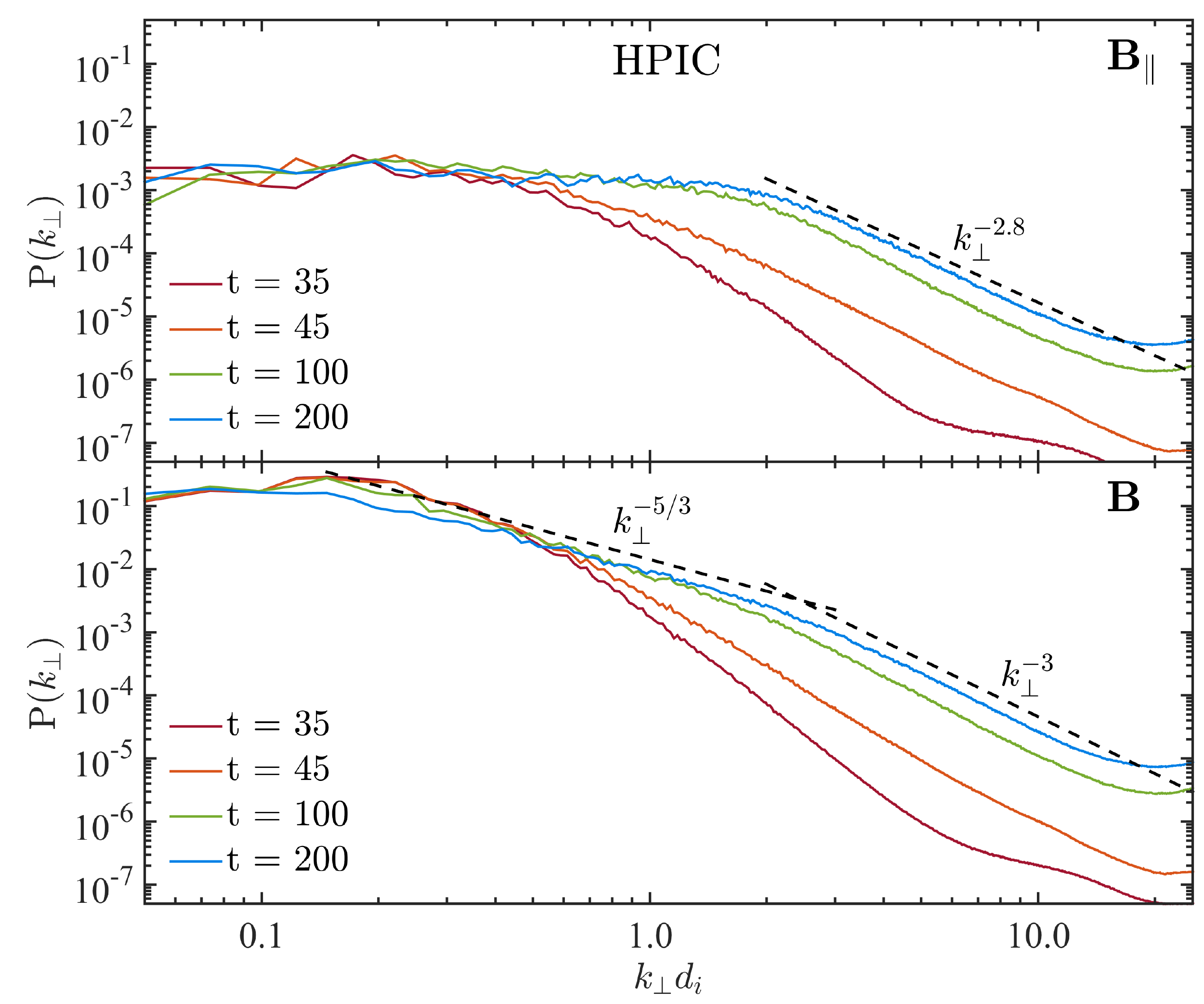}
\caption{Power spectra of the parallel (top) and total 
  (bottom) magnetic fluctuations for the HPIC run
  before (red) and after (orange) the first magnetic 
  reconnection events occur, at an intermediate time (green) and when the
  quasi-steady state is reached (light blue). Characteristic power laws 
  are drawn as a reference.}
\label{fig:SpectraHPIC}
\end{figure}
\begin{figure}[t!]
\includegraphics[width=0.47\textwidth]{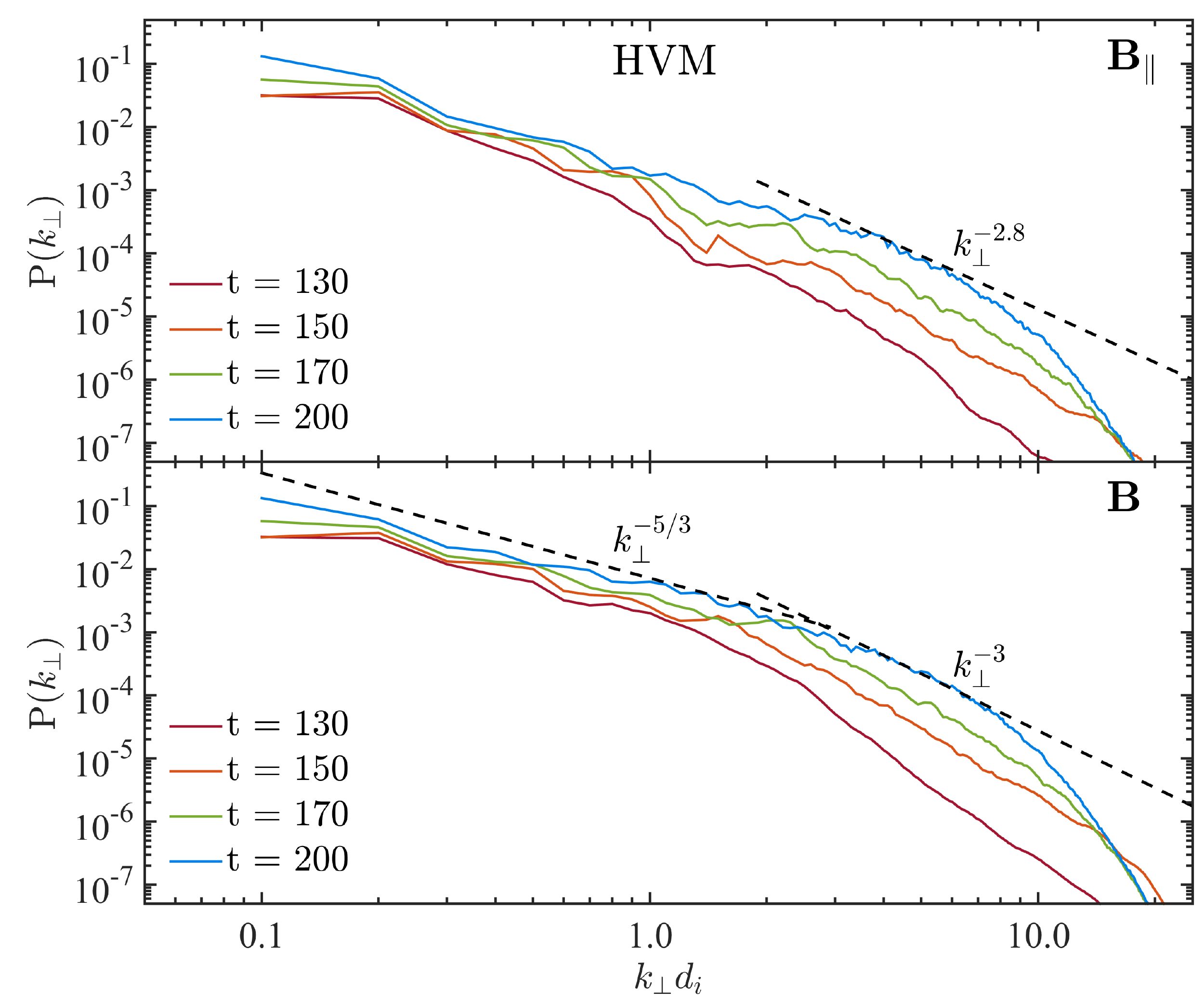}
\caption{The same as in Fig.~\ref{fig:SpectraHPIC}, but for the HVM
  run.}
\label{fig:SpectraHVM}
\end{figure}

We now focus on the effects of magnetic reconnection processes on 
the spectral properties. 
We first look at the power spectra of the parallel magnetic fluctuations,
$\Bv_\|$, for the HPIC simulation (Fig.~\ref{fig:SpectraHPIC}, top).  
At $t = 35$, before reconnection has occurred, no clear
power law is observed, even at MHD scales. 
Soon after the first reconnection event ($t \sim 45$), a power 
law develops at sub-ion 
scales, with a spectral index of $\sim -2.8$. This 
value is typically observed in density and parallel magnetic field 
in 2D simulations \citep{FranciAPJL2015}, regardless of the plasma
beta \citep{FranciAPJ2016}. At later times, the level of fluctuations in the
kinetic range gradually increases, keeping the same slope. 

A similar evolution is observed for the total magnetic
fluctuations, $\Bv$ (Fig.~\ref{fig:SpectraHPIC}, bottom), with two
main differences: i) the asymptotic slope at sub-ion scales is steeper,
around $-3$, and attained gradually, later than the first
reconnection event; ii) the MHD part of the spectrum continues to
flatten slowly, until a $-5/3$ power-law develops, much later. 
This behavior is consistent with a picture where reconnection
drives a kinetic-scale turbulent cascade, which is better
appreciated in the $\Bv_\|$ fluctuations (as well as in the 
density fluctuations, not shown here) and, only later, the direct
cascade from larger scales bring its contribution to the total magnetic power spectrum,
due to the Alfv\'enic-like
$\Bv_{\perp}$ component.  The time required for the
formation of a stable and extended power law in the kinetic-range,
once reconnection has started occurring at $t \sim 40$, is $\sim 5$
inverse ion gyro-frequency, i.e., shorter than the eddy turnover time, 
$\tau_{\mathrm{inj}}^{\mathrm{nl}} \gtrsim 20$. Moreover,
Fig.~\ref{fig:TimeEvolution_RmsMaxJ}d shows that the inverse
reconnection rate of the first event, 
$\tau^{\mathrm{SG1}}_\mathrm{rec}$, is indeed smaller than the
eddy turnover time estimated at both the injection scale,
$\tau^{\mathrm{inj}}_\mathrm{nl}$, and at kinetic scales,
$\tau^{\mathrm{kin}}_\mathrm{nl}$, at the time when the reconnected
flux starts increasing.  The latter is given by
$\tau^{\mathrm{kin}}_\mathrm{nl} = (k_{\perp}^\mathrm{inj}
\mathbf{u}_e^\mathrm{inj})^{-1} \approx ({k_{\perp}^\mathrm{inj}}^2
\Bv^\mathrm{inj})^{-1}$, where $\mathbf{u}_e$ is the electron bulk
velocity and we chose the scale $k_{\perp} \, d_i = 2$,
which will later correspond to the spectral break.
The comparison of $\tau^{\mathrm{SG1}}_\mathrm{rec}$ with
$\tau^{\mathrm{inj}}_\mathrm{nl}$ and
$\tau^{\mathrm{kin}}_\mathrm{nl}$ indicates that reconnection is
indeed very efficient in transferring energy at kinetic scales, faster
than a direct cascade from the injection scale, causing a strong
and rapid decrease of $\tau^{\mathrm{kin}}_\mathrm{nl}$
(cf. Fig.~\ref{fig:TimeEvolution_RmsMaxJ}d) in correspondance with
the first local maximum of $\max(|\Jv|)$
(cf. Fig.~\ref{fig:TimeEvolution_RmsMaxJ}b) and the sudden increase of
$\Phi$ (cf. Fig.~\ref{fig:TimeEvolution_RmsMaxJ}c).

\begin{figure}[t]
\includegraphics[width=0.49\textwidth]{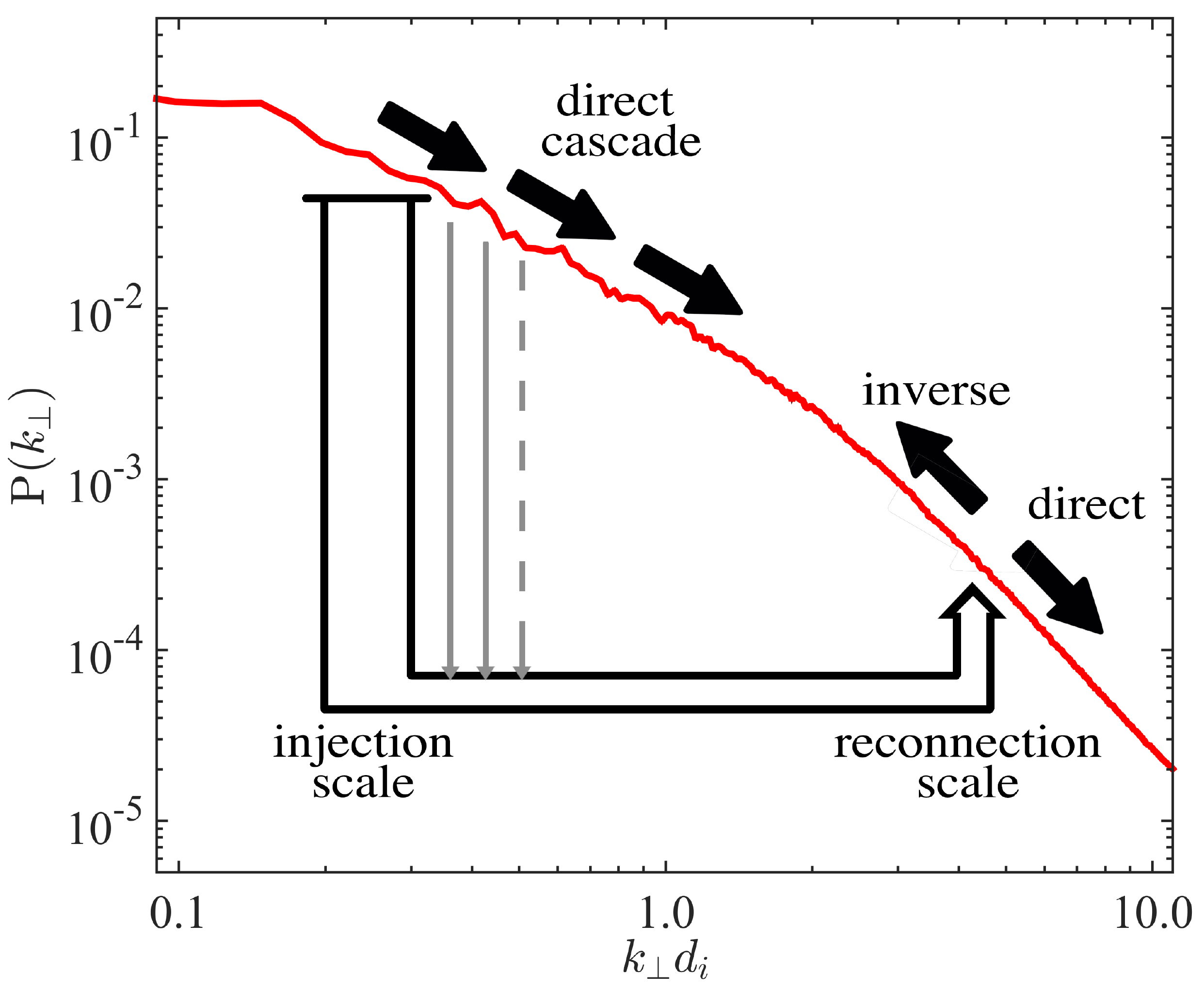}
\caption{Schematic view of the development of the turbulent cascade as a
combination of two mechanisms: i) direct cascade through local
transfers in Fourier space, ii) injection of energy from large-scale
vortices directly into small-scale structures via reconnection,
local in real space but non-local in Fourier space.}
\label{fig:SchematicCascade}
\end{figure}

This analysis, together with the evidence that a kinetic cascade forms
rapidly and despite the absence of a Kolmogorov-like cascade at large
scales, suggests that the kinetic spectrum is not a simple
extension of the MHD spectrum through a ``classic'' cascade involving
only local interactions in $k$-space.  Energy
is directly injected at small scales via non-local interactions in
Fourier space mediated by magnetic reconnection occurring in strong
and thin current sheets, whose width is of the order of the ion
scales~\citep[e.g.][]{FranciAPJ2016}. 
Reconnection events produce ion-scale magnetic islands, which
can merge, initiating an inverse cascade towards larger scales, or can
start a transfer of energy towards smaller scales via a direct
cascade.  This channel for the generation of the magnetic field
spectrum at sub-ion scales is sketched in Fig.~\ref{fig:SchematicCascade}.  
Concurrently, although on larger time scales, a direct turbulent
cascade develops from the largest scales, generating eddies of smaller
and smaller sizes, which interact and form many other current sheets,
injecting additional energy at ion
scales. This mechanism can be appreciated by looking at the evolution
of magnetic fluctuations

Let's now consider the magnetic field spectra of the forced HVM
simulation (Fig.~\ref{fig:SpectraHVM}). Here, the path to fully
developed turbulence is reversed compared to the HPIC decaying case.
Initially, the energy at large scales grows slowly, due to the
external forcing, and develops into a Kolmogorov-like cascade at
$t\sim 120$, while no significant power is present at kinetic scales
yet.  Later, once the large-scale fluctuations reach roughly the same
level as in the HPIC case, reconnection starts occurring (at
$t_{\mathrm{rec}}^{\mathrm{HVM}} \sim 135$, cf. Fig.~\ref{fig:2DPlots}
g-j), and a power-law spectrum forms also at kinetic scales, with the
same spectral index of $-2.8$ in $\Bv_\|$.  Finally, a stationary
regime is reached, characterized by a double power-law behavior, in
agreement with the HPIC simulation~\citep{CerriJPP2017}.

\section{Conclusions} 

In the present work, we have provided the first numerical evidence that a
sub-ion-scale cascade in collisionless plasmas can develop
independently from a Kolmogorov-like cascade at MHD scales, triggered
by magnetic reconnection.  This new picture of the turbulent dynamics
across the ion break has been achieved by analyzing two
high-resolution hybrid simulations, which employ different methods to
drive the turbulence and different numerical methodologies to simulate
the system evolution, a Lagrangian hybrid particle-in-cell 
and an Eulerian hybrid Vlasov-Maxwell approach.

In HPIC, an extended power law in the spectrum of the parallel
magnetic fluctuations forms early at sub-ion scales, without a
well-developed turbulent spectrum at MHD scales. Only later, a
Kolmogorov-like cascade for the (total) magnetic fluctuations
gradually develops in the MHD range, due to the contribution of the
perpendicular components.  In HVM, conversely, the same power law at
kinetic scales is achieved after a Kolmogorov-like MHD cascade is
established, only as soon as the first reconnection event has
occurred.  In both cases, a fully-developed turbulence state is
achieved, in which both the MHD and the sub-ion spectral slopes are
quasi-stationary and attain the values of $-5/3$ and $-3$,
respectively.  Furthermore, the kinetic range exhibits the same
properties in the two cases (comparable level of parallel and
perpendicular fluctuations, see~\citet{CerriJPP2017}), despite the
quite different MHD-range behavior: in HPIC, the cascade is carried by
perpendicular fluctuations, while in HVM the parallel component
dominates.  

The correspondance between the onset of magnetic reconnection events
and the formation of a stable power-law spectrum at kinetic scales,
together with the fact that the inverse reconnection rate is initially
much shorter than the eddy turnover time around ion scales, is a clear
evidence that in both simulations the former acts as a trigger for the
latter.  The present analysis does not allow us to determine whether
or not reconnection still remains the main energy source feeding the
small-scale turbulence also once a stationary state is reached.
Nevertheless, the interaction between large-scale eddies keeps driving
the formation of many, randomly distributed, current sheets. When
these undergo reconnection, their width, intensity, and reconnection
rate are still of the same order of the early events.  Although we do
not quantitatively evaluate the competition between reconnection and
the standard wave-wave interaction as energy transfer mechanisms, we
conjecture that the former is likely the preferred/fastest path for
energy injection at ion scales, based on the fact that: i) the local
maxima of $\max(|\Jv|)$, directly linked to reconnection events,
exhibit approximately the same intensity from $t_\mathrm{rec}$ on,
indicating that strong current sheets keep forming and disrupting, ii)
$\tau_\mathrm{nl}^\mathrm{kin}$ rapidly decreases as soon as reconnection 
begins, adjusting and settling to an asymptotic value 
$\tau_\mathrm{nl}^\mathrm{kin} \lesssim 5 \,\Omega_i^{-1}$, 
which is comparable to the inverse reconnection rate $\tau_\mathrm{rec}$, 
and iii) once formed, the power law at kinetic scales
 is well maintained and the spectra only grow in amplitude
until the quasi-stationary state is reached, indicating that the
number of reconnecting current sheets increases until a balance
between formation and disruption is achieved.  We suggest thus that
any theory of the turbulence cascade down the ion scales should
carefully take into account the role of the magnetic reconnection
which should not be seen only as the location where dissipative
effects are dominant.

\begin{acknowledgments}
The authors wish to acknowledge valuable discussions with D. Burgess,
T. Horbury, C. H. K. Chen, and J. E. Stawarz.  S.C.C. and F.C. thank
F.~Rincon for the implementation of the external forcing and
C. Cavazzoni (CINECA, Italy) for his essential
contribution to the HVM code parallelization and performances. L.F. is
funded by Fondazione Cassa di Risparmio di Firenze, through the
project ``Giovani Ricercatori Protagonisti''. L.M. was funded by the
UK STFC grant ST/N000692/1.  P.H. acknowledges GACR grant 15-10057S.
The authors acknowledge PRACE for awarding access to resource
Cartesius based in the Netherlands at SURFsara through the DECI-13
(Distributed European Computing Initiative) call (project HybTurb3D),
CINECA for awarding access to HPC resources under the ISCRA initiative
(grants HP10BUUOJM, HP10BEANCY, HP10B2DRR4, HP10CGW8SW, HP10C04BTP),
and the Max Planck Computing and Data Facility (MPCDF) in Garching
(Germany).
\end{acknowledgments}

\bibliographystyle{yahapj}


\end{document}